\documentclass[11pt,a4paper]{article}
\usepackage{jcappub}
\pdfoutput=1
\usepackage{graphicx}
\usepackage{adjustbox}  
\usepackage{booktabs}
\usepackage[T1]{fontenc}
\usepackage[utf8]{inputenc}
\usepackage{url}
\usepackage{appendix}
\usepackage{amsmath}
\usepackage{mathtools}
\usepackage{subfig}
\usepackage{epstopdf}
\usepackage{listings}
\usepackage{verbatim}   
\usepackage{tabularx,ragged2e}
\usepackage{booktabs}

\def\bea{\begin{eqnarray}}
\def\eea{\end{eqnarray}}
\def\be{\begin{equation}}
\def\ee{\end{equation}}
\def\fr{\frac}

\def\la{\label}
\def\be{\begin{equation}}
\def\ee{\end{equation}}

\def\le{\left}
\def\ri{\right}

\def\LCDM{$\Lambda$CDM\,}

\def\lfs{\lambda_{fs}}

\title{\boldmath The virial mode $k_v$  approach to  Structure Formation with Warm Dark Matter}

\author[a,b,1]{Axel de la Macorra \note{Corresponding author}}
\author[c,d]{and Jorge Mastache}

\affiliation[a]{Instituto de F\'isica, Universidad Nacional Autonoma de M\'exico, Circuito de la Investigaci\'on 
Cient\'ifica Ciudad Universitaria, 04510, CDMX, M\'exico.}
\affiliation[b]{Instituto de Ciencias del Cosmos, University of Barcelona, ICCUB, Barcelona 08028, Spain.}
\affiliation[c]{Mesoamerican Centre for Theoretical Physics, Universidad Aut\'{o}noma de Chiapas,  Carretera Zapata Km. 4, Real del Bosque, 29040, Tuxtla Guti\'{e}rrez, Chiapas, M\'{e}xico.}
\affiliation[d]{Consejo Nacional de Ciencia y Tecnolog\'ia, Av. Insurgentes Sur 1582, Colonia Cr\'edito
Constructor, Del. Benito Ju\'arez, 03940, Ciudad de M\'exico, M\'exico} 

\emailAdd{macorra@fisica.unam.mx, a.macorra@gmail.com}
\emailAdd{jhmastache@mctp.mx}

\abstract{
The small scale structure opens a window to constrain the dynamical properties of Dark Matter. Here we study the clustering of warm dark matter (WDM) in a semi-analytical approach and compared the linear power spectrum of WDM with cold dark matter (CDM) employing a new transfer function $T_v(a,k)$ in terms of the viral wave number  $k_v=2\pi/\lambda_v$  corresponding  to  a structure with  a viral radius $r_v=\lambda_v/2$,  half the size of the free streaming scale radius  $r_v=r_{fs}/2=\lambda_{fs}/4$.  The virial mass $M_v$ contained in this structure corresponds to the lightest structure formed for a WDM particle becoming non-relativistic at the scale factor $a_{nr}$ with the corresponding $\lambda_{fs}$. The  viral transfer function  $T_v(a,k)=[1+ \left(k/k_v \right)^{\beta_v }]^{\gamma_v}$ is given in terms of the viral mode $k_v$ and two constant parameters $\beta_v$ and $\gamma_v$. We compare $T_v(a,k)$ with the Boltzmann code  CLASS for WDM in the mass range 1-10 keV and we obtain the constraint $\beta_v\gamma_v=-18$ with  $\nu=1.020 \pm 0.025$. In the standard approach the transfer function is given by $T(a,k)=[1+\left(\alpha\, k \right)^{\beta}]^{\gamma}$  \cite{Viel:2005qj} 
where $\alpha$ encodes the dynamical properties of  WDM and must be numerically adjusted by means of a Boltzmann code.  In contrast, in our  viral approach the physical quantity $k_v$ is  simply given in terms of the free streaming scale $\lambda_{fs}$ and can be analytically determined. 
Our viral proposal has a good agreement with CLASS and improves slightly  the results from the standard  transfer function.
To conclude, we have proposed a  new physically motivated transfer function $T_v(a,k)$   where the properties of WDM are encoded in the viral wave number $k_v$,  
is straightforward to determine and improves the prediction of WDM clustering properties. 
}


\begin{document}
\maketitle
\flushbottom

\section{Introduction}\label{sec:introduction}

The standard $\Lambda$CDM model  has been a very successful model to describe our Universe and is consistent with
cosmological observational evidence such as the cosmic microwave background (CMB) anisotropies \cite{Aghanim:2018eyx}, galaxy redshift surveys \cite{Abbott:2016ktf}, type Ia Supernovae \cite{Betoule:2014frx} reach to the conclusion that the content of the Universe is composed of 69\% dark energy driving the accelerated expansion of the Universe, 31\%  matter whose clustering feature influence the large scale structure formation, corresponding to 27\% Dark Matter (DM) and the remaining 4\% is baryonic matter. 

The nature of dark matter has received a great deal of attention in the last decade, due in part of the missing satellites in the universe, and the amount of structure at different scales and redshifts puts strong constraints on the nature of dark matter. In recent times the large surveys such as SDSS-IV \cite{Alam:2020sor}  and in the near future, DESI \cite{Levi:2019ggs} in the near future will have an important impact in determining the properties of dark energy (DE) and dark matter (DM).
Despite the efforts in both particle physics and cosmology, the nature and composition of the DM are still unknown. Candidates for DM can be classified according to its velocity dispersion, $v_{nr}$, when the DM particle become non-relativistic (given by the scale factor $a_{\rm nr}$). 
For thermal relics one can relate the mass of the WDM particle to $a_{\rm nr}$. In this case DM  is cold   for mass  larger than  $m_{\rm cdm} \sim \mathcal{O}$(MeV). 
This kind of DM particles stop being relativistic and start clustering object at early times. DM with a mass around $m_{\rm wdm} \sim \mathcal{O}$(keV) are known to be warm, WDM, whose main attribute is that its dispersion velocity wipes out some density concentrations of matter and, therefore, induce a cut-off scale into the mass halo function \cite{Archidiacono:2013dua}. The amount of energy density today $\Omega_{\rm dmo}$ along with the scale factor when the WDM particle becomes non-relativistic $a_{nr}$ are crucial for determining the properties of the large scale structure of the universe. For instance, the dispersion velocity of DM particles wipes out density concentrations of matter and, therefore, induce a cut-off scale in the mass halo function \cite{Archidiacono:2013dua}. Most DM candidates have a smooth evolution of the DM velocity, however, phase transitions in the underlying particle physics model for dark matter shows that an  abrupt transition to a non-relativistic limit is plausible \cite{ delaMacorra:2009yb,Mastache:2019bxu}.

The large velocity dispersion of DM at early stages on the evolution of the Universe tends to suppress gravitational clustering at small scales and conciliates with what it is observed.  Cosmological N-body simulations of the \LCDM model predict the number of satellite galaxies in Milky Way-like galaxies is smaller than the expected, the so-called missing satellite problem \cite{Kazantzidis:2003hb, BoylanKolchin:2011de, Klypin:1999uc} and high concentrations of DM in the innermost regions of galaxies (cusp-core problem \cite{Mastache:2011cn,Marsh:2015wka}). Baryon physics also pursued the solutions to this problem by integrating star formation and halo evolution in the galaxy, however, the discussion is still in progress \cite{Garrison-Kimmel:2017zes, Sawala:2015cdf, Pawlowski:2015qta}.  
The details of the suppression at small scales depend on the DM particle nature, which takes us to connect the DM models and astrophysical observations. It can be determined by the parametrization of the transfer function $T_X(a,k)=(P_X/P_{cdm})^{1/2}$ in terms of the power spectrum of $X$ DM particles and \LCDM model, where $T_X$ is scale and time-dependent. Is has been conventional to compare the $X$ and CDM models at the epoch when the amplitude of the fluctuations of model $X$ is half the size of \LCDM,  i.e. $T_X(a,k)^2=1/2$. Different ansatzes have been proposed to parametrize the transfer function $T_X(a,k)$ and the parameters involved must be numerically fitted using numerical Boltzmann codes \cite{Viel:2005qj,Viel:2013fqw}.  

Here, we will present a new approach to structure formation where the virial radius places a dominant role in structure formation. This new approach is physically motivated and is consistent with previous works in WDM structure formation and it allows for an understanding of the suppression of small scale structure in terms of the virial mass and radius.  We introduce a new analytical  transfer function, physically motivated,  that reproduces the clustering of large classes non-thermal DM models and preserving the connection with the physics and nature of the DM.  

We present and compare the standard and our virial transfer function in section \ref{STF} and  section \ref{VTF} , and the free streaming scale in Section \ref{ssec:freestreaming}, the dark matter analysis and finally the conclusion in section \ref{sec:conclusion}.

 \section{Warm Dark Matter}\label{WDM}
 
 The clustering properties of Dark Matter  have a direct impact on the number of halo as a function of mass and redshift 
and  can be contrasted with  several observational large scale structure experiment as \cite{Alam:2020sor}. 
The velocity dispersion of DM particles plays a crucial roll in structure formation. While DM particles are still relativistic, primordial density fluctuations are suppressed due to the velocity dispersion of DM particles and  inhibit the formation of structure below the free streaming scale $\lambda_{fs}$ \cite{Bode:2000gq}.  
Here,  we will present two approaches  to extract cosmological clustering properties of warm dark matter (WDM)
 in terms of  Transfer Function $T( a, k )$ defined as the quotient of the linear matter perturbations between WDM and CDM.
We refer to the standard approach the work of  M. Viel et al. \cite{Viel:2005qj,Viel:2013apy}  presented in section \ref{STF}
while our virial approach is given in section  \ref{VTF}.
 
Throughout this paper, we  adopt Planck 2018 cosmological parameters \cite{Aghanim:2018eyx} in a flat Universe with $\omega_{dmo} = 0.12$, and $\omega_{bo} = 0.02237$ as the CDM matter and baryonic with $\omega_i=\Omega_i h^2$  (with $ i=dm,b$) and $h = 0.6736$ the Hubble constant in units of 100 km\,s$^{-1}$Mpc$^{-1}$,  $z_{\rm reio} = 7.67$  the reionization redshift , $n_s = 0.965$  the tilt of the primordial power spectrum and $\ln(10^{10} A_s) = 3.044$ with $A_s$ the amplitude of primordial fluctuations.

 \subsection{Transfer Function: Standard Approach}\la{STF}

We will now  present the standard approach to extract cosmological clustering properties of warm dark matter (WDM)  using the Transfer Function defined as the quotient of the linear matter perturbations between WDM and CDM. 
The velocity dispersion of dark matter particles inhibits the formation of structure and the main parameter to account
for this dispersion is  the scale factor  when these particles become non-relativist  given by $a_{nr}$ 
\cite{Bode:2000gq} and recently in \cite{Mastache:2019bxu}. 
The evolution of the linear  energy density fluctuations are conveniently calculated  by publicly available Boltzmann codes, e.g. CAMB \cite{Lewis:2002ah} and CLASS  \cite{Lesgourgues:2011re}, giving a power spectrum of matter-energy density fluctuations $P_{lin}(a,k)$. By comparing the power spectrum for different types of WDM particles one can infer the properties of a new WDM model without the need to implement the new model in the Boltzmann codes.
The parametrization of the matter power spectrum for different WDM models has been presented in \cite{Viel:2005qj, Viel:2013fqw},  where the properties of WDM models can be studied employing the transfer function $T(a,k)$, defined as 
 \be
 T( a, k ) = \left[ \fr{P _ { \mathrm { lin } }^{ \mathrm { \rm wdm }}(a,k) }{P_ { \mathrm { lin }}^{ \mathrm { cdm } }(a,k)}  \right]^{1/2}
\label{eq:trasfer_mps}
\ee 
 by comparing the power spectrum in the WDM model with CDM. The clustering properties of WDM are conveniently  determined by the scale $k_{1/2}$  mode, where the power spectrum of the WDM model is suppressed by $50\%$ compared to a \LCDM model, i.e.  
 \be
 \le [T( a, k_{1/2} )\ri]^2=\fr{1}{2}
\ee
and  it is a convenient  reference point, in Table \ref{tab:khalf_all} we show some $k_{1/2}$ values for different WDM models.
The standard transfer function for thermal WDM particle  is given by 
\cite{Viel:2013fqw, Viel:2005qj,Bode:2000gq, Colin:2000dn, Hansen:2001zv}
\begin{equation}
  T  ( k )  = \left[ 1 + (\alpha k ) ^ {\beta} \right]^{\gamma}.
 \label{eq:transfer_viel}
 \end{equation}
The clustering properties of the dark matter model is contained in $\alpha$   while  $\beta$ and $\gamma$ are constant parameters to be fitted from numerical simulations.  The quantity $\alpha$  has been estimated  in \cite{Viel:2013fqw, Viel:2005qj}
\be
\alpha= 0.049 \le( \fr{m_{\rm wdm}}{1 keV} \ri)^{1.11}   \le( \fr{\Omega_{\rm wdm}}{0.25} \ri)^{0.11}  \le( \fr{h}{0.7} \ri)^{1.22}  h^{-1} Mpc.  
\la{alpha}\ee
We can take  $\beta$ and $\gamma$ as independent parameters or  constrain them to be proportional to a single parameter $\mu$ as in 
\cite{Viel:2005qj} where they find
\be\la{vielparam}
    \beta=2\mu,\;\;\;\;\;\;\;\;\;  \gamma=-5/\mu,\;\;\;\;\;\;\;\;\;   \mu = 1.12
\ee
with $\beta\gamma=-10$.

\subsection{Analysis with  $k_\alpha$  mode}

Alternatively to $k_{1/2}$, we can define the mode $k_\alpha\equiv 1/\alpha$ rendering $T(k_\alpha)=2^\gamma$. 
From Eq.\eqref{eq:transfer_viel} we see that the half  mode is proportional to $1/\alpha$ 
\be
k_{1/2} = \fr{1}{\alpha} \le[\le(\fr{1}{\sqrt{2}}\ri)^{1/\gamma}-1\ri]^{1/\beta}=\fr{1}{\alpha}\; \xi_V
\la{kV1}\ee
with a proportionality constant 
\be
\xi_V \equiv \le[\le(\fr{1}{\sqrt{2}}\ri)^{1/\gamma}-1\ri]^{1/\beta}.
\la{xi1}
\ee
For the values given in eq.(\ref{vielparam}) we obtain $\xi_V \simeq 0.36$. We define the mode
\be
k_{\alpha}\equiv \fr{1}{\alpha} = \fr{k_{1/2}}{\xi_V}
\la{ka1}
\ee 
and $k_{\alpha} > k_{1/2}$. We see that for $k=k_{\alpha}$  the transfer function takes the  constant value 
\be
T ( k_\alpha )= 2^{\gamma} 
\la{Tx1}
\ee
with $T(k_\alpha) \simeq 0.045$ for $\gamma=-5/\mu, \mu=1.12$.

The ratio of the transfer function $T _ { \mathrm { X} } ( k_{\alpha})/T _ { \mathrm { X} } ( k_{1/2})\simeq 1/10$  but $k_\alpha/ k_{1/2} \simeq 3$. Since $k_\alpha$  is larger than $k_{1/2}$ it corresponds to a less linear mode with  a smaller transfer function  $T _ { \mathrm { X} } ( k_{\alpha})$ indicating a larger deviation from  \LCDM model.
The information  using  $T _ { \mathrm { X} } ( k_{1/2}^V)=1/2$ or $T _ { \mathrm { X} } ( k_{\alpha})= 2^\gamma$   and $k_{\alpha}$ is  the same and $k_\alpha$ provides  an alternative mode  to compare WDM models using the transfer function.

\section{The Virial Model}\la{VTF}

We propose to determine the transfer function $T _ v ( k_v )$ as a function of the viral mode $k_v$, defined in terms of the  free streaming mode  with  $k_v=2\pi/\lambda_v$ with $r_v=\lambda_v/2$ half the radius of as structure with a size of the free streaming scale $\lfs$.
We refer to  he transfer function $T _ v ( k_v )$ as the virial approach.
The free streaming scale $\lambda_{fs}$  and mass $M_{fs}$ are
\begin{eqnarray}
  \lambda_{fs} (t)&\equiv& \int_0^{t} \frac{v(a)  dt'}{a(t')}, \;\;\;\;\;\;\;\;\;\;  M_{fs}=\le(\fr{4\pi\,  r_{fs}^3}{3}\ri)    \rho_o, 
\la{Mfs}
\end{eqnarray}
with $r_{fs}=\lambda_{fs}/2$ and $\rho_o$ present matter content of the universe. The  virial mode, radius  and viral mass are given by
\be
k_v=2\pi/\lambda_v,\;\;\;\;\;\;\; r_v=\fr{\lambda_{v}}{2} =\fr{\lambda_{fs}}{4},\;\;\;\;\;\;\;\;\;  M_v=\le(4\pi\, r_v^3/3)\ri) \rho_v. 
 \ee
The mass contained in the virialized  structure is conserved and the virial mass is $ M_v =  M_{fs}$ and since $r_v$ is half the size of $r_{fs}$ the  density of the virial sphere is  $\rho_v= 8 \rho_0$.  Of course, a halo density profile  such as NFW \cite{Navarro:1995iw, Navarro:1996gj} for example, would be more appropriate  to determine the energy density inside this structure but  this lies beyond the scope of this work. 
We take the transfer function for the virialized mode $k_v$  as a two-parameter linear transfer function $T _ v ( k_v )$ as a function of the viral mode $k_v$,  similar as the Standard Transfer, 
\begin{equation}
  T _ v ( k ) =  \left[ 1 + \left( \frac{k}{ k_{v}} \right) ^ { \beta_v} \right] ^ { \gamma_v } 
\la{TF}\end{equation} 
with $\beta_v$ and $\gamma_v$  constant parameters.   We determine the values of these parameters in 
the next section (\ref{VTF}), where we find that $\beta_v$ and $\gamma_v$ are correlated leaving $T _ v ( k )$ with only one free parameter.

\begin{figure}
\centering 
     \includegraphics[width=0.75\textwidth]{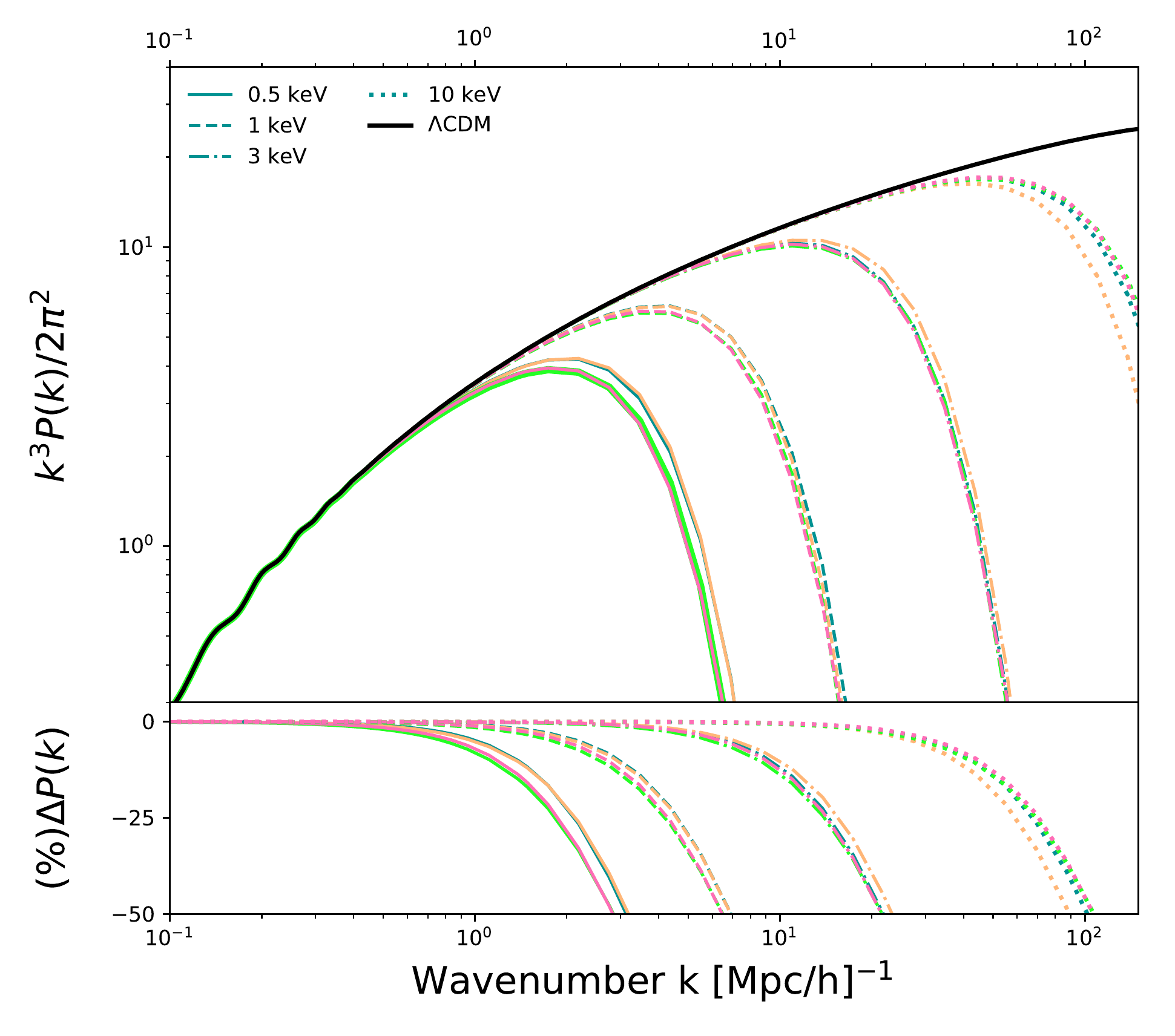} 
    \caption{\footnotesize{ {\bf Top panel}. Plots of linear dimensionless matter power spectra for the CDM (black solid line) and WDM of different masses, 0.5, 1, 3, and 10 keV, plotted as straight, dashed, dot-dashed and dotted lines, respectively. Dark green color lines are those obtained from CLASS. Orange color lines is the power spectrum is given from the parametrization, Eq.\eqref{eq:transfer_viel}, pink and light green lines are the virial parametrization form Eqs. \eqref{eq:transfer_2p}  and \eqref{eq:transfer_1p}, respectively. We show the percentage difference, $\Delta P(k)$, with respect the CDM matter ower specrum.  }}  \label{fig:mps}
\end{figure}

 \subsection{Transfer Function: Virial Approach}

\begin{figure}[t]
\centering
    \includegraphics[width=0.75\textwidth]{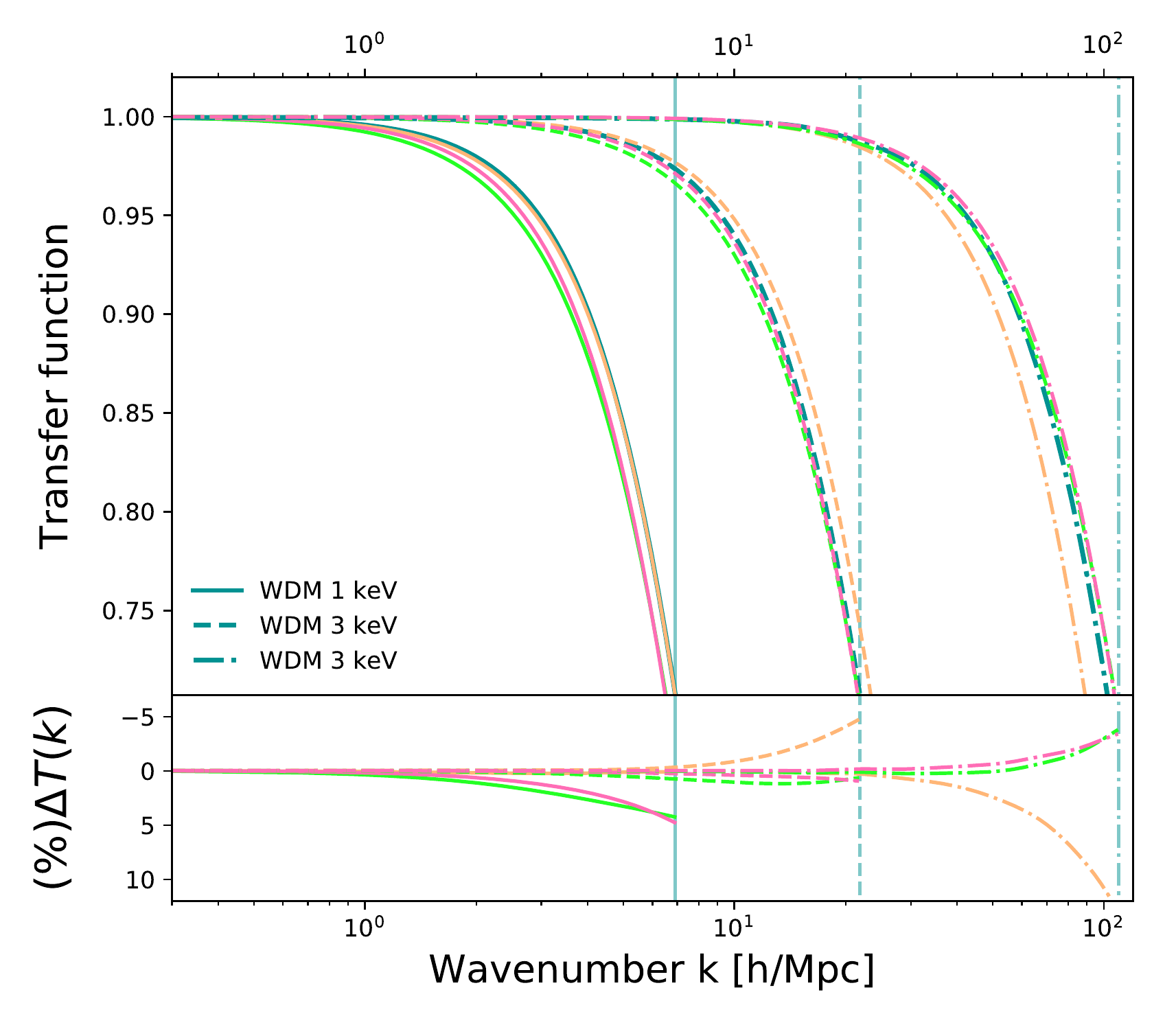}
    \caption{\footnotesize{
        Transfer function for WDM. Solid, dashed, and dot-dashed lines represent WDM masses of 1, 3, and 10 keV, respectively. Dark green color are transfer functions from CLASS code, orange color are the transfer functions from Viel  et al. \cite{Viel:2005qj} Eq.\eqref{eq:transfer_viel}, pink lines are the virial transfer functions with 2 free parameter (Eq.\ref{eq:transfer_2p}), and green lines is the transfer function with 1 free parameter (Eq.\ref{eq:transfer_1p}). }}
    \label{fig:transfer_fuct}
\end{figure}

Let us now determine compare the transfer function $T _ v ( k )$ with Boltzmann codes for different WDM models.
\begin{equation}
  T _ v ( k ) =  \left[ 1 + \left( \frac{k}{ k_{v}} \right) ^ { \beta_v} \right] ^ { \gamma_v } 
\end{equation} 
with $\beta_v$ and $\gamma_v$  constant parameters. All the physical properties of WDM are imprinted in the virial mode $k_v$ which is directly  related to the  free-streaming scale $\lambda_{fs}$ given in eq.\eqref{lfsT}  and it is not
adjusted by the Boltzmann codes, contrary to the $\alpha$ parameter in  eq.(\ref{alpha}) in section \ref{STF}.
The dependence of $k$ in the transfer function in eq.\eqref{TF} is given by
\be
\fr{k}{k_v}=  \fr{r_v}{ r}
\ee
The scale  $k=k_v = 2k_{fs}$ corresponds to a radius $r_v$ which is half the size  of  the free streaming scale radius $r_{fs}=\lambda_{fs}/2$. 
We therefore  have  virial radius and $k$ mode 
\be
r_v \equiv  \fr{r_{fs}}{2}  =\fr{ \lambda_{fs}}{4},\;\;\;\;\;\;\;\;\;\;\;\; k_v= 2k _{fs} = \fr{4\pi}{\lambda_{fs}}.
\la{kv}
\ee
The dependence of the transfer function on the mass of the WDM particle is in the virial mode, $k_{v}$, and not in the $\alpha$ parameters of the transfer function as in eq.\eqref{TF}). The mode $k_v$ can be easily determined since it is proportional to the free streaming scale (cf. eq.(\ref{lfs0})) and does not require  to be fitted using a numerical Boltzmann code. To determine the  values of  $\beta_v$ and $\gamma_v$ we compare our transfer virial function with WDM models from Class code \cite{Lesgourgues:2011re}, see Fig.\ref{fig:transfer_fuct}.

We found numerically convenient to define the parameters  $\beta_v$ and $\gamma_v$  as $ \beta_v=2\nu$ and $\gamma_v=-\zeta/\nu$  and we take $\zeta$ and $\nu$ as the independent parameters and we obtain
for WDM masses in the range (1-10)keV using CLASS, using  the result for
$a_{nr}$ given in eq.(\ref{eq:anrao}) to compute  $\lambda_{fs}$ and $k_v=4\pi/\lfs$,  
\begin{equation}
	T =   \left(1 + \left(\frac{k}{ k_{v}} \right)^{2\nu}\right)^{-\zeta/\nu}.
\label{eq:transfer_2p}
\end{equation} 
\begin{equation}
\beta_v=2\nu, \;\;\;\;\; \gamma_v=-\zeta/\nu,\;\;\;\;\;\;\;\;\;\;\;\;	\nu = 1.0253 \pm 0.0232, \;\;\;\;\;\zeta = 9.17 \pm 0.19.
\la{virial2}\end{equation}
giving a constraint $\beta_v\gamma_v = -2\zeta=18.34 \pm 0.38$.
We also consider the case with  a single free parameter, as in the standard approach, by relating $\beta_v$ and $\gamma_v$. 
Contrary to the ansatz in the standard transfer function approach, here we  choose to define  $\beta_v=2\nu$ and $\gamma_v=-9/\nu$ giving a constraint $\gamma_v\beta_v= -18$ 
\begin{equation}
T = \left(1 + \left(\frac{k}{ k_{v}} \right)^{2\nu}\right)^{-9/\nu}
\label{eq:transfer_1p}
\end{equation} 
with
\begin{equation}
\beta_v=2\nu,\;\;\;\;\;\;\;  \gamma_v=-9/\nu, \;\;\;\;\;\;\;\;\;\;\;\; 	\nu = 1.020 \pm 0.025.
\la{virial1}\end{equation}
Notice that the value of $\nu$ in the one parameter case in eq.(\ref{virial1}) is consistent with the two parameters in eq.(\ref{virial2})
and allowing us to work with the  transfer function in eq.(\ref{eq:transfer_1p}).

In Table \ref{tab:khalf_all} and Fig.\ref{fig:transfer_fuct} and Fig.\ref{fig:mps} we show how good the virial transfer functions is by computing the $k_{1/2}$ for different models, for the standard transfer function, $k_{1/2}^{\rm viel}$, and also the virial approach with 2 free parameters, eq.\eqref{eq:transfer_2p} denoted as $k_{1/2}^{2p}$ and with 1 free parameter, eq.\eqref{eq:transfer_1p} and denoted as $k_{1/2}
^{1p}$. The numbers is parenthesis are the percentage difference in comparison with the scale directly obtained from the numerical solution from CLASS, $k_{1/2}^{\rm class}$.  Class code
 \cite{Lesgourgues:2011re}, see Fig.\ref{fig:transfer_fuct}. We show in table \ref{tab:k_virial} the values of $k_v,\,r_v,\, M_v$  and $a_{nr} $ or different WDM masses in the range 1-10keV   in our viral approach.

\begin{table*}[t]
\centering     
\small{
\begin{tabular}{r|rrrr|r|r}
\hline
Mass  & $a_{nr}$ \;\;\;\;\; & $k_{v}$ \;\;& $k_{1/2}^{\rm v\,1p}$\;\;\;\;\; & $k_{1/2}^{\rm v\,2p}$\;\;\;\;\;\;\; & $k_{1/2}^{\rm viel}$\;\;\;\;\;\;\; & $k_{1/2}^{\rm class}$\;\;\;  \\
\hline
\hline
 1 keV & $1.20\times10^{-7}$ & 31.58 & 6.53 (5.60\%) & 6.53 (5.60\%) & 6.92 \,\,(0.11\%) & 6.91   \\
 2 keV & $4.97\times10^{-8}$ & 69.78 & 14.40 (4.40\%) & 14.47 (4.88\%) & 14.93 \,\,(8.25\%) & 13.79    \\
 3 keV & $3.18\times10^{-8}$ & 104.98 & 21.64 (1.03\%) & 21.73 (0.58\%) & 23.37 \,\,(6.88\%) & 21.86    \\
 4 keV & $1.98\times10^{-8}$ & 161.82 & 33.40 (3.60\%) & 33.40 (3.60\%) & 32.21 \,\,(7.03\%) & 34.65    \\
 5 keV & $1.39\times10^{-8}$ & 223.48 & 46.26 (6.05\%) & 46.26 (6.05\%) & 41.31 \,\,(5.29\%) & 43.62    \\
 6 keV & $1.09\times10^{-8}$ & 280.32 & 58.00 (5.61\%) & 58.00 (5.61\%) & 50.41 \,\,(8.20\%) & 54.91    \\
 7 keV & $8.83\times10^{-9}$ & 339.04 & 70.13 (1.44\%) & 70.13 (1.44\%) & 59.86 (13.41\%) & 69.13    \\
 8 keV & $7.39\times10^{-9}$ & 399.62 & 82.53 (5.17\%) & 82.53 (5.17\%) & 69.50 (20.15\%) & 87.03    \\
 9 keV & $6.31\times10^{-9}$ & 461.86 & 95.39 (9.60\%) & 95.39 (9.60\%) & 79.24 \,\,(8.95\%) & 87.03    \\
 10 keV & $5.56\times10^{-9}$ & 519.30 & 107.29 (2.08\%) & 107.29 (2.08\%) & 89.13 (18.65\%) & 109.57 \\
\hline
\end{tabular}}
\caption{ Viral wavenumbers $k_{v}$ and $k_{1/2}$ for different transfer functions  in hMpc$^{-1}$ units. Values for $k_{1/2}^{\rm v\, 1p}$ and $k_{1/2}^{\rm v\,2p}$ are obtained by the transfer function given by eq.\eqref{eq:transfer_1p} and eq.\eqref{eq:transfer_2p}, respectively, while  $k_{1/2}^{\rm viel}$ is obtained from eq.\eqref{eq:transfer_viel}.
The values of $a_{nr}$ and  $k_{1/2}^{\rm class}$, 2nd and last column respectively, were obtained from the transfer function given by the code CLASS.} 
\label{tab:khalf_all}
\end{table*}

\subsection{Analysis with the Viral mode $k_v$}

For $k=k_v$ the transfer function takes the value $T _ { \mathrm { X} } ( k )=(2) ^{\gamma_v }$ , independently of the value of $\beta_v$,  and can be used to compare different DM models as $k_\alpha$ (c.f. eq.(\ref{ka1})) given in the standard approach. From eq.(\ref{TF}) the half mode $k^v_{1/2}$  (i.e. $T^2 _v (k^v_{1/2} ) =1/2$) is given by  
\be
k_{1/2}^v = k_v\, \xi_v
\la{kv2}\ee
with
\be
\xi_v \equiv \le[\le(\fr{1}{\sqrt{2}}\ri)^{1/\gamma_v}-1\ri]^{1/\beta_v}
\la{xi1v}\ee
with a constant value given by $\xi_v \simeq 0.21$ using  eq.\eqref{virial1}.

\begin{figure}[t]
  \centering
    \includegraphics[width=1\textwidth]{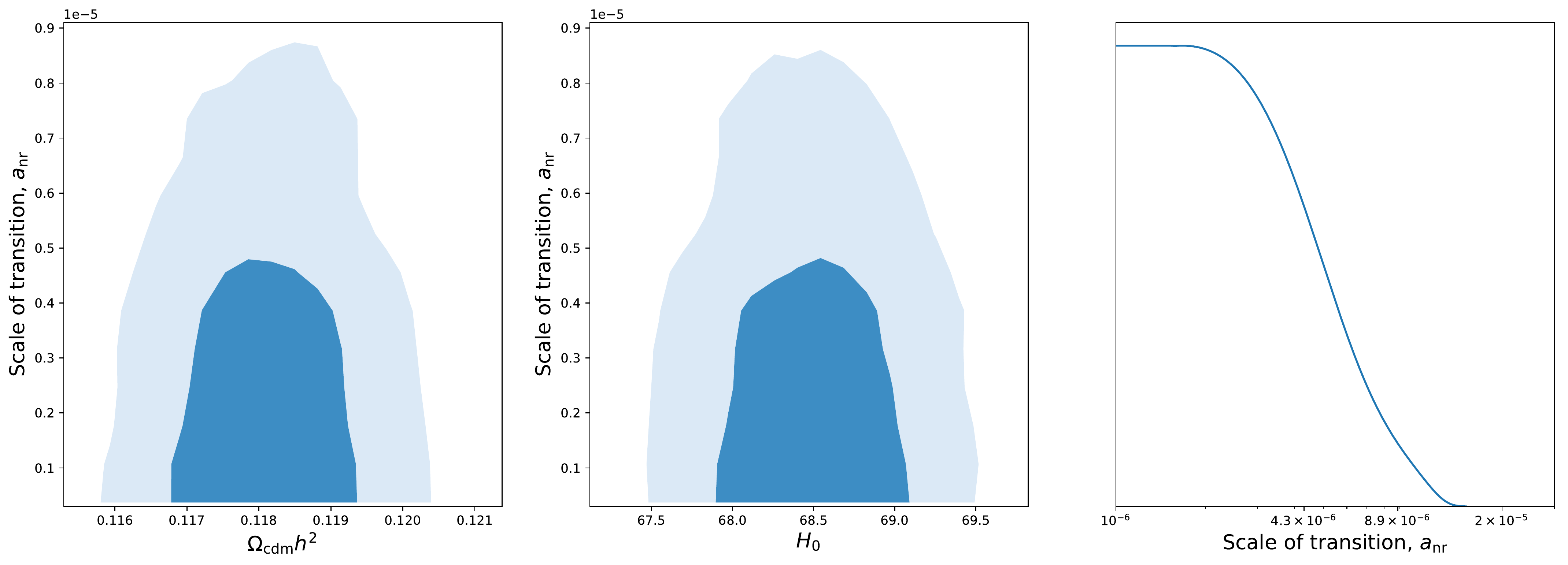}
    \caption{\footnotesize{We show recent constrains \cite{Mastache:2019gui} to $a_{nr}$ from posterior probabilities at 68\% and 95\% confidence level in the parameter space $a_{nr}-\Omega_{\rm cdm}$ (left), $a_{nr}-H_0$ (middle), and the marginalized posterior distribution of $a_{nr}$ (right) using Planck EE, TE, and TT high-$\ell$ spectra \cite{Aghanim:2018eyx} combined with Wigglez matter power spectrum \cite{Parkinson:2012vd} observations. At $1\sigma$ confidence level, the lower value is given by $a_{nr} < 4.3\times 10^{-6}$. }}
    \label{fig:likelihood}
\end{figure}

\begin{table*}[t]
    \centering     
	\begin{tabular}{l|cccc}
	\hline
        Mass &	 $a_{nr}$ &	 $r_v$ &	 $k_v$ &	  $M_v$   \\
       \hline
       \hline
        1 keV &	 $1.20\times 10^{-7}$ &	 99.8 &	 31.5 &	 $1.32\times 10^{9}$  \\
        2 keV &	 $4.76\times 10^{-8}$ &	 43.3 &	 72.6 &	 $1.07\times 10^{8}$  \\
        3 keV &	 $2.77\times 10^{-8}$ &	 26.4 &	 118.8 &	 $2.45\times 10^{7}$  \\
        4 keV &	 $1.89\times 10^{-8}$ &	 18.6 &	 168.7 &	 $8.56\times 10^{6}$  \\
        5 keV &	 $1.40\times 10^{-8}$ &	 14.2 &	 221.5 &	 $3.78\times 10^{6}$  \\
        6 keV &	 $1.10\times 10^{-8}$ &	 11.3 &	 276.9 &	 $1.93\times 10^{6}$  \\
        7 keV &	 $8.96\times 10^{-9}$ &	 9.4 &	 334.6 &	 $1.10\times 10^{6}$  \\
        8 keV &	 $7.50\times 10^{-9}$ &	 8.0 &	 394.2 &	 $6.70\times 10^{5}$  \\
        9 keV &	 $6.41\times 10^{-9}$ &	 6.9 &	 455.6 &	 $4.34\times 10^{5}$  \\
        10 keV & $5.57\times 10^{-9}$ &	 6.1 &	 518.7 &	 $2.94\times 10^{5}$  \\
        \hline
	\end{tabular}
   \caption{ Virial quantities for different WDM masses, with $a_{nr}$ the scale factor of the WDM transition given by Eq.\eqref{eq:anrao}, $r_v$ the virial radius in kpc/h is given from Eq.\eqref{kv}, the wave number $k_{v}$ in Mpc$^{-1}$h, and the virial mass $M_v$ in $M_\odot/h^3$.}
   \label{tab:k_virial}
\end{table*}

\subsection{Comparing the two approaches}

We have presented and alternative approach to the Transfer function. In the standard transfer function approach in section \ref{STF}
 the quantity   $\alpha$  given in eq.(\ref{alpha})  contains the properties of DM model such as its mass (which can be related to the scale factor when it becomes non-relativistic  $a_{nr}$ for thermal WDM particles), however the numerical values $\alpha$ have been numerically fitted using a Boltzmann code \cite{Viel:2005qj}.  In our virial approach, all model dependence  is encoded in the free-streaming scale $\lfs$,  which is well defined function and is not numerically fitted.

We plot  the Matter Power Spectrum and the transfer functions  in Fig.\ref{fig:mps} in the mass range (1 - 10) keV for WDM. Notice that the standard error for all three parameters are small, specially for $\alpha_v$ and $\beta_v$. The value of $\beta_v=2.04$ is actually close to previous works  that obtained $\beta= 2.24$ \cite{Viel:2005qj}, however we find a significantly difference with  $\gamma_v = -2.68$  compared to $\gamma_ = -4.46$. The difference in these parameters is because we take a different reference mode $k_v$ instead of $k_\alpha=1/\alpha$ as in Viel et al. \cite{Viel:2005qj} with similar proportionality constants $\xi_v=0.21$ and $\xi_V=0.36$.

Our  transfer function  is physically motivated and is determined in  terms of the virial mode $k_v$ which has a clear interpretation in terms of the free streaming scale and in contrast with
the fitted $\alpha$ parameter  (c.f.  eq.\eqref{alpha}) in the standard approach  \cite{Viel:2013fqw,Viel:2005qj}. Our transfer function  has an excellent agreement once we compare with the Boltzmann numerical code, CLASS and it out performs the standard Transfer function in eq.\eqref{eq:transfer_viel}   for massive particles in the range 1-10keV . 


\section{Free-streaming scale and Virial radius $r_v$}\label{ssec:freestreaming}

\begin{figure}[t]
  \centering
    \includegraphics[width=0.75\textwidth]{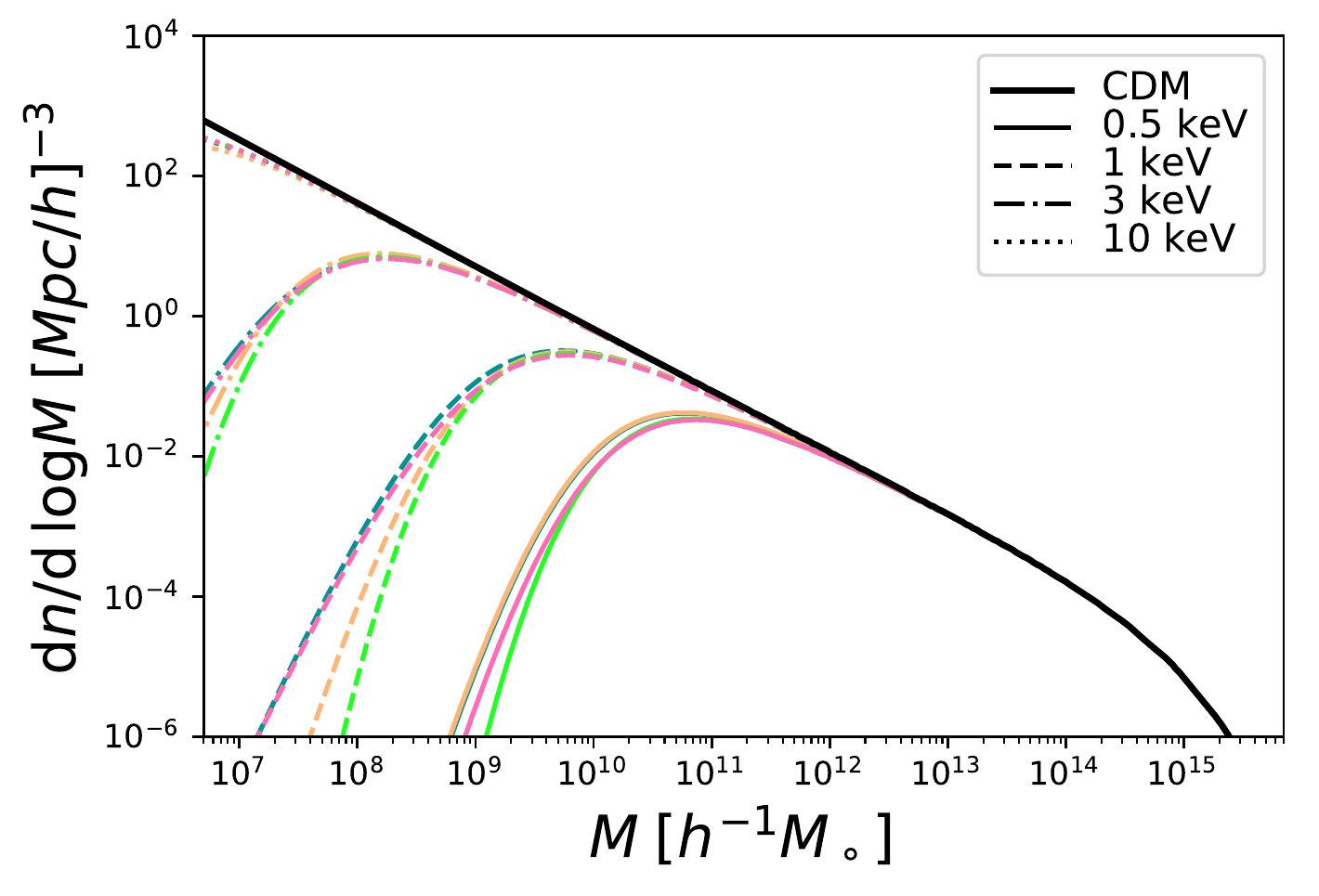}
    \caption{\footnotesize{
        Plot of the halo mass function. The black solid line is the CDM model; Color straight, dashed, dot-dashed and dotted line styles represent 0.5, 1, 3, and 10 keV WDM masses respectively. Color lines represent the halo mass function from different matter power spectrum. Dark green are the lines from CLASS, pink and light green lines are the virial parametrization form Eqs. \eqref{eq:transfer_2p}  and \eqref{eq:transfer_1p}, respectively. And orange color lines is the standard parametrization, Eq.\eqref{eq:transfer_viel}.  }}
    \label{fig:pressschechter}
\end{figure}

The thermal velocities of the dark matter particles have a direct influence on structure formation. While DM particles are still relativistic, primordial density fluctuations are suppressed on scales of order the Hubble horizon at that time. This is called the free-streaming scale and depends on the time when a massive particle becomes non-relativistic $t_{nr}$. It is defined by
\begin{eqnarray}
\la{lfs0}
  \lambda_{fs} (t)&\equiv& \int_0^{t} \frac{v(a)  dt'}{a(t')} 
\end{eqnarray}
The corresponding free-streaming scale  mode $k_{fs}$  and a mass $M_{fs} $ contained in sphere of radius $\lambda_{fs}/2$ are given by
\be \label{eq:free_streaming_scale}
k_{fs}=\fr{2\pi}{\lambda_{fs}}, \;\;\;\;\;\;\;\;\; M_{fs} = \fr{4\pi}{3} \le(\fr{\lambda_{fs}}{2}\ri)^3 \rho_{mo}. 
\ee
and in terms of the viral mode
\be
 r_v=\fr{r_{fs}}{2}=\fr{\lambda_{fs}}{4}, \;\;\;\;\;\;\;\;\; M_v=\le(\fr{4\pi\, r_v^3}{3}\ri).    
\la{rmv}\ee
with $\rho_v=8\rho_{mo}$. For mass-scales $M \lesssim M_{v}$ the free-streaming of particles erases all peaks in the initial density field therefore the number of structures below this mass scale should be significantly reduced in numbers. We show this behavior in Fig. \ref{fig:pressschechter}, where we compare CDM and WDM mass functions. However, it is the virialized scale that accounts for the mass contained within the structure to be formed.

\subsection{Free Streaming scale}

It is conventional to report  the free streaming scale  assuming  a  relativistic regime  with $v_{\rm wdm}=c=1$ for $a<a_{nr}$ and  following a  non-relativistic regime ($a>a_{nr}$) 
with $v_{\rm wdm}=a_{nr}/a$.  With these choices of $v$  one gets the  free streaming scale  
\bea
 \lambda_{fs} (t_{eq}) &\simeq& \int_0^{t_{nr}} \frac{c dt}{a(t)}   + \int_{t_{nr}} ^{t_{eq}}\frac{v_{\rm wdm} dt}{a(t)} , \\
    &= &\fr{2t_{nr}}{a_{nr}} \le[1+ \ln\le(\fr{a_{eq}}{a_{nr}}\ri) \ri]
\la{lfs1}\eea
to be compared with of Eq.\eqref{lfsT} or its limit eq.\eqref{fs2a}. 
However, we prefer to follow our previous approach   \cite{Mastache:2019bxu, Mastache:2019gui} and define the time when a particle stops being relativistic,  at scale factor $a_{nr}$, when  $p^2 = m^2$ with $p$ the momentum of the particles.  The velocity in an expanding universe evolves as
\begin{equation}
v(a)  = \frac{ (a_{nr}/a)}{\sqrt{1 + (a_{nr}/a)^2}},
\label{veldm} \end{equation}
which gives a velocity $v^2_{\rm wdm}(a_{nr}) = 1/2$. Eq.(\ref{veldm}) describes the exact velocity evolution of a decoupled massive particle. The transition between relativistic to non-relativistic is smooth and continuous, see \cite{Mastache:2019bxu} for a generalize transition. This evolution is general and valid for any massive decoupled particles (WDM, CDM or massive  neutrinos). Using eq.(\ref{veldm}) in eq.(\ref{lfs1}) we get  \cite{Mastache:2019bxu})
 \be
\lambda_{fs} (a_{eq}) = \fr{2t_{nr}}{a_{nr}}\ln\left [\frac{a_{eq}}{a_{nr}}+\sqrt{1 + \left(\frac{a_{eq}}{a_{nr}}\right)^2} \right],  
\la{lfsT}\ee
which gives  free streaming scale without approximation. However, for presentation purposes  let assume that 
 $a_{nr}\ll a_{eq}$  and then
\be
\lambda_{fs} (a_{eq})\simeq  \fr{2t_{nr}}{a_{nr} }\le[\ln(2) + \ln\left (\frac{a_{eq}}{a_{nr}} \right)\ri].
 \la{fs2a} 
 \ee
Here, we use in our calculations eq.\eqref{lfsT} but  for presentation purposes we take the approximations in eq.(\ref{fs2a}).
We express  $2t_{nr}=1/H_{nr} $ and   $H_{nr}=  (a_{eq}/a_{nr})^2(a_o/a_{eq})^{3/2} H_o$  to  obtain
 \bea
\fr{2t_{nr}}{a_{nr}}  &=& 0.011  \left(\frac{1+z_{eq}}{1+3411}\right)^{1/2} \left( \frac{a_{nr}}{2.77\times10^{-8}} \right) \frac{\rm Mpc}{\rm h}
 \eea
where $a_{nr}$ has been previously computed in \cite{Mastache:2019bxu},
\be
\fr{a_{nr}}{a_o} =  2.77  \times 10^{-8}   \le(\fr{\Omega_{dmo}h^2}{0.120}\ri)^{1/3} \le(\fr{3\,keV}{m_{\rm wdm}}\ri)^{4/3}  \, ,
\la{eq:anrao}\ee
where we consider the DM particle as a  fermion with $g_f=7/4$ degrees of freedom. With a simple analytic approach for a massive particles characterized by having a non-negligible thermodynamic velocity dispersion \cite{Bode:2000gq, Colin:2000dn, Hansen:2001zv, Viel:2005qj, Dodelson:1993je, Dolgov:2000ew, Asaka:2006nq, Shi:1998km, Abazajian:2001nj, Kusenko:2006rh, Petraki:2007gq, Merle:2015oja, Konig:2016dzg} we can compute the fluid approximation for the perturbation equations for any massive particles,   given the analytic solution for the energy density evolution we were able to reproduce the most appealing feature of WDM, the cut-off in the matter power spectrum \cite{Colin:2000dn, Hansen:2001zv,Viel:2005qj, Viel:2013fqw}.  The percentage difference between the numerical value obtained from Boltzmann equations and the analytic one of $a_{nr}$ is on average 3\% in a mass range 1-10 keV.

In Table \ref{tab:k_virial} we show different values of $a_{nr}$ and in Fig. \ref{fig:likelihood} we show the lower limits that constrains $a_{nr}$ from the fluid approximation \cite{Mastache:2019gui} applied to WDM.
The free-streaming scale $\lambda_{fs}$ and mass in the limit $a_{nr}/a_{eq} \ll 1$ take the following values
\begin{eqnarray}      
      \lambda_{fs} &\simeq& 97.0 \left( 1 +\le(\fr{1}{8.82}\ri) \left( \frac{1+3411}{1+z_{eq}} \right) \left( \frac{2.77\times10^{-8}}{a_{nr}} \right) \right) \frac{\rm kpc}{\rm h}     
\la{lfssol} \end{eqnarray}
with  $\lambda_{fs}=108.2\; \rm kpc/\rm h$ for $z_{eq}=3411,\;a_{nr}=2.77\times10^{-8}$, and a contained mass of
\begin{eqnarray}
    M_{fs}^{\rm wdm} = 2.45\times10^7\le( \frac{\Omega_{dmo}h^2}{0.12}\ri)\left( \frac{\lambda_{fs}}{108.2\; \rm kpc} \right)^3 M_\odot.
\la{massfs}
\nonumber \end{eqnarray}
The virialized structure  has a   mass $M_v=M_{fs}$, however the energy density  in the virialized  structure is eight times larger than in $M_{fs}$,  and a  radius  $r_v=r_{fs}/2=\lambda_{fs}/4$  with
 \bea
 \la{rvsol}
 r_v &=& 24.25\left( 1 + \le(\fr{1}{8.82}\ri)\left( \frac{1+3411}{1+z_{eq}} \right) \left( \frac{2.77\times10^{-8}}{a_{nr}} \right) \right) \frac{\rm kpc}{\rm h}  \\
 &=&  26.4\; \le(\fr{\lambda_{fs}}{108.2\; {\rm kpc/h} }\ri) \frac{\rm kpc}{\rm h}.
 \eea

\subsection{Halo Mass Function}\label{ssec:preschechter}

The comoving number density of collapsed structure is computed using the Press-Schechter formalism \cite{Press:1973iz}. With the linear matter power spectrum obtained from the transfer function, eq.\eqref{eq:trasfer_mps}, we compute the halo mass function having mass range $M$ to $M + dM$  as
\begin{equation}
  \frac{dn}{d\log M} = \frac{1}{2} \frac{\overline{\rho}}{M} \mathcal{F}(\nu) \frac{d \log \sigma^2}{ d \log M}
\end{equation}
where $n$ is the number density of haloes, $M$ the halo mass and the peak-height of perturbations is given by $\nu = \frac{\delta_c^2(z) }{\sigma^2(M)}$, where  $\delta_c = 1.686$ is the critical overdensity required for a structure to collapse in a dark matter halo in the $\Lambda$CDM cosmology. The average matter density is $\overline{\rho}$. The corresponding variance of the smoothed density fluctuation in a sphere of radius $R$ enclosing a mass
$M$ is $\sigma^2(M)$, can be computed from the following integrals
\begin{equation} \label{eq:sigma_8}
  	\sigma^2(M) = \int_0^\infty dk \frac{k^2 P_{\rm lin}(k)}{2\pi^2}| W(kR) |^2.
\end{equation}
Here we will use the sharp-k window function $W(x) = \Theta( 1 - kR)$, with $\Theta$ being a Heaviside step function (smoothes the large scale mass distribution to a continuous density field), and $R = (3cM/4\pi \overline{\rho})^{1/3}$, where the value of $c = 2.5$ is proved to be best for cases similar as the WDM \cite{Benson:2012su}. Finally, for the mass function, $\mathcal{F}(\nu)$, we adopt \cite{Bond:1990iw} that is giving as
\begin{equation}
   \mathcal{F}(\nu) = A\left( 1 + \frac{1}{\nu^{\prime p}} \right)\sqrt{ \frac{\nu^\prime}{2\pi} } e^{-\nu^\prime/2}
\end{equation}\label{eq:pressschechter}
with $\nu^{\prime} = 0.707\nu$, $p = 0.3$, and $A = 0.322$ determined from the integral constraint $\int f(\nu)d\nu = 1$. 

Using the matter power spectrum using the transfer function with Eq.\eqref{eq:trasfer_mps}, and using the virial approach Eqd.\eqref{eq:transfer_1p} and \eqref{eq:transfer_2p}. We compute the halo mass function, see Fig.\ref{fig:pressschechter}, and compare it with the mass function obtained from CLASS numerical solutions to the matter power spectrum and the same from Viel transfer function, Eq.\eqref{eq:transfer_viel}.

\section{Summary and Conclusion} \label{sec:conclusion}

We studied the clustering properties of WDM particles and we present  the viral approach, where the transfer function 
depends on the viral  wave number  $k_v=4\pi/\lfs$, and we  compare it to the standard approach. The velocity dispersion of WDM at early stages on the evolution of the universe suppress gravitational clustering and may conciliates observations of
 the number of small scales galaxies. 
 The details of the suppression depend on the properties of  the  DM particles, however a key ingredient
is the time when these particles become non-relativist given by the scale factor $a_{nr}$. 

We have presented here a new approach to structure formation where the virial radius $r_v$ places a dominant role.  The viral mode $k_v$ is defined in terms of the free streaming scale $\lfs$ and  depends thus directly on $a_{nr}$. 
This new approach is physically motivated, is consistent with previous works in WDM structure formation, and allows for an understanding  the suppression of small scale structure in terms of the virial mass and radius, $M_v$ and $r_v$ respectively. The transfer function $T(a,k)$ is determined in terms of the virial mode $k_v$ which has a clear interpretation in terms of the free streaming scale and is easily calculated, contrasting with the $\alpha$ parameter of eq.\eqref{alpha} \cite{Viel:2005qj,Viel:2013fqw} which requires a numerical fit.

From  Table (\ref{tab:khalf_all}) we see that our virial transfer function performs better than the standard transfer function for WDM with masses in the range (1-10) keV.  The standard transfer function has up to two constant parameters ($\beta,\gamma$) which can be reduced to only one $\mu$  with $\beta=2\mu, \;\gamma= - 5/\mu$ with $\mu=1.12$ supplemented by the $\alpha$  containing the  relevant  clustering parameters 
of the WDM model  in eq.(\ref{alpha}) and has been fitted to give the correct value of $k_{1/2}$ for different WDM models. Since $\alpha$ has been numerically fitted we consider it as a free parameter.
 Therefore the standard transfer function has up to three parameters and can be reduced to two parameters by taking the fitted constraint $\gamma \beta =-10$.
On the other hand our viral Transfer function parametrization in eq.\eqref{eq:transfer_2p} has only two parameters $\beta,\gamma$. As in the Standard Transfer function we can reduce a parameter  by taking $\gamma_v \beta_v$  constant. Doing so, we found  $\beta_v=2\nu,\, \gamma_v=-9/\nu$  with $\nu=1.09$, and $\gamma_v\,\beta_v=-18$. Taking $\beta_v$ and $\gamma_v$ independent  we obtained $\beta_v =  2.05 \pm 0.04$, $\gamma_v= -8.94 \pm 0.10$ with $\gamma_v\,\beta_v=-18.32$  at the central values.  
Clearly  the one  and two  parameter  Transfer Function in the viral approach are consistent.
We obtained for a WDM thermal particle  that becomes non-relativistic  at  $a_{nr} = 2.77 \times 10^{-8}$, corresponding to a 3 keV mass,  a viral radius $r_v=26.4$M$_{pc}$ with wave number  $k_v=21.64$ in the viral approach and a  $k_{1/2}=21.86$ in the standard approach, corresponding to a structure with a
contained mass of $\mathcal{O} (10^7) \; {\rm M}_{\odot}$.

We  have shown that our  virial parametrization of the transfer function is physically motivated and  has an excellent agreement  with numerical results  from Boltzmann CLASS codes shown in Table (\ref{tab:khalf_all}) and Fig.(\ref{fig:transfer_fuct}) and Fig.(\ref{fig:mps})  and outperforming slightly the standard transfer function 
\cite{Viel:2005qj, Bode:2000gq}  given in eq.\eqref{eq:transfer_viel} for masses in the range 1-10 keV.  
The parameter $\alpha$ in the Standard Transfer function must be fitted numerically using Boltzmann code, while in our approach
the physical parameter   $k_v=2k_{fs}$  is determined in  terms of the free streaming mode $\lfs$ an easily computed quantity.

\acknowledgments
AM  acknowledges partial support from  Project IN103518 PAPIIT-UNAM, PASPA DGAPA-UNAM and CONACyT. JM acknowledges Catedras-CONACYT financial support and MCTP/UNACH as the hosting institution of the Catedras program.

\bibliographystyle{unsrtnat}
\bibliography{WDM_virial}

\end{document}